\documentclass[letterpaper,twocolumn,10pt]{article}
\usepackage{usenix,epsfig}
\usepackage{todonotes}
\usepackage{booktabs}
\usepackage[symbol]{footmisc}
\usepackage{graphicx}
\usepackage{svg}
\usepackage{pdfpages}
\usepackage{fancyhdr}
\usepackage{flushend}

\begin{document}

%don't want date printed
\date{}

%make title bold and 14 pt font (Latex default is non-bold, 16 pt)
\title{\Large \bf GDPArrrrr: Using Privacy Laws to Steal Identities}

\author{
{\rm James Pavur}\thanks{Corresponding Author.~Email Address:~first.last@cs.ox.ac.uk~or first.last@pavursec.com. NB: All views are those of the authors and not their respective organizations.}\\
DPhil Researcher\\
Oxford University
\and
{\rm Casey Knerr}\\
Security Consultant\\
Dionach LTD
}

\maketitle

% Use the following at camera-ready time to suppress page numbers.
% Comment it out when you first submit the paper for review.
%\thispagestyle{empty}

\subsection*{Abstract}
The General Data Protection Regulation (GDPR) has become a touchstone model for modern privacy law, in part because it empowers consumers with unprecedented control over the use of their personal information. However, this same power may be susceptible to abuse by malicious attackers. In this paper, we consider how legal ambiguity surrounding the ``Right of Access'' process may be abused by social engineers. This hypothesis is tested through an adversarial case study of more than 150 businesses. We find that many organizations fail to employ adequate safeguards against Right of Access abuse and thus risk exposing sensitive information to unauthorized third parties. This information varied in sensitivity from simple public records to Social Security Numbers and account passwords. These findings suggest a critical need to improve the implementation of the subject access request process. To this end, we propose possible remediations which may be appropriate for further consideration by government, industry and individuals.

\section{Introduction}
The General Data Protection Regulation (GDPR) represented a sea change in the way European residents could control, restrict, and understand the use of their personal information. As arguably the most consequential attempt to regulate data security to date, GDPR's ambition has caused millions of organizations to reflect on and even revise their cyber-security and data collection practices. One particularly notable feature of GDPR is the power it gives to individual consumers to act as the first line of defense against data abuse. Under GDPR, European residents have the right to request, review, amend, and delete, personal information which organizations store about them.

In this paper, we consider the practical implementation of this right, with a particular focus on mechanisms to prevent its abuse to steal sensitive information about a third party. We find that GDPR itself provides little guidance on best practices and, more broadly, that little attention has been paid to the possibility of request abuse for the purpose of data theft. This lacuna is contextualized through a real-world experiments in which simulated fraudulent GDPR requests are sent to more than 150 organizations.

Our experimental findings demonstrate that many organizations fail to adequately verify the originating identity of right of access requests. The result is that social engineers can abuse right of access requests as a scalable attack vector for acquiring deeply sensitive information about individuals. We suggest several mitigations which may help remediate this situation, both in the status-quo context of GDPR and in the context of emerging next-generation privacy laws.

\section{Background and Motivation}

GDPR came into effect on May 25, 2018, as an ambitious replacement to decades-old European privacy legislation. The law outlines new requirements for organizations handling personal data, including serious fines for non-conformance. The largest offences could incur fines of up to €20 million or 4\% of a company's annual revenue - whichever is greater (Ch. 8, Art. 83)~\cite{noauthor_general_2016}. Companies in violation of the law have already begun to feel these consequences. For example, Google has been fined €50 million by the French Commission nationale de l'informatique et des libertés (CNIL) for its handling of personal data for advertisements~\cite{cellan-jones_google_2019}. Additionally, Marriott International and British Airways have been fined over £99 million and £183 million, respectively, by the UK Information Commissioner's Office (ICO) for breaches resulting in the compromise of customer data~\cite{uk_information_commissioners_office_statement:_2019, uk_information_commissioners_office_intention_2019}. However, some surveys suggest that many small business owners are still uncertain that they are fully compliant with the law~\cite{gdpr.eu_gdpr_2019}.

A fundamental component of GDPR is the ``right of access'' to one's personal data (Ch. 3, Sec. 2). To exercise this right, residents of the European Union are permitted to send subject access requests (SARs) to almost any organization. The recipient of the SAR is then permitted one month to respond, ideally with a copy of all the personal data the organization holds on the sender.

According to the text of the GPDR, an organisation is permitted to employ ``all reasonable measures to verify the identity of a data subject who requests access'' (Rec. 64). As more specific details are not readily available, organizations may be inclined to interpret this requirement on a case-by-case basis. Further, the law specifies that organizations are not permitted to collect data for the sole purpose of verifying identities in response to future SARs (Rec. 64). This severely restricts the spectrum of viable identity validation measures - especially for data-brokers and other organizations without direct consumer interaction.

An organization not otherwise exempt from GDPR requirements is only permitted to refuse to comply with an SAR for one of two reasons (Ch. 3, Art. 12). The first refusal condition is met if an SAR is deemed ``excessive'' in frequency. This applies to individuals who send the same or similar SARs within a very short period of time. The second refusal condition is met if an SAR is deemed ``manifestly unfounded.'' The UK ICO lists six example reasons why this condition may apply to an SAR under their ``Guide to the GDPR"~\cite{uk_information_commissioners_office_right_2019}. These reasons include using GDPR requests as a component of blackmail, to harass or disrupt and organization, to target particular employees, or to waste organizational resources. No direct association is made between the term ``manifestly unfounded'' and the threat of fraud. Moreover, the burden of proof appears to fall to the organization, with the UK ICO stating that organizations ``are responsible for demonstrating that [the request] is manifestly unfounded''~\cite{uk_information_commissioners_office_right_2019}.

These reasons relate to malicious intent on the part of the sender but do not discuss the possibility of fraud directly - focusing instead on the abuse of GDPR requests to waste organizational resources.

An organization refusing to comply with an SAR for one of the reasons above is still required to respond to the sender with the reason for refusal~\cite{uk_information_commissioners_office_right_2019}. According to the UK ICO, the response should also remind the sender of their rights to complain to regulation authorities or to seek legal action against the refusing organisation. It it therefore fairly risky to fail to provide data in response to an SAR, even for a valid purpose. For an organization concerned about receiving large fines or damaging their reputation, it may be less dangerous to comply with potentially invalid requests than to risk challenging legitimate ones. 

Calculations such as the above may leave organizations vulnerable to social engineering attacks which target the SAR process. Often, skilled social engineers will attempt to create an artificial sense of urgency or fear to pressure victims into divulging sensitive information. However, under GDPR both time pressure and the threat of fines already exist as a natural effect of the regulation. Additionally, as the GDPR provides a clear mechanism for gaining access to sensitive data, there is no need for an attacker to invent a sophisticated pretense for the initial request for information beyond a simple reference to GDPR rights. In short, the right of access process appears intuitively well-suited to abuse by social engineers and ambiguity surrounding both identity verification and request denial grounds in GDPR further bolsters this possibility.

\section{Experimental Design}
The objective of our experiment was to determine if an unsophisticated attacker might acquire sensitive personal information about an individual through malicious subject access requests. Casey Knerr, the second author of this paper, consented to play the role of ``victim'' in the experiment. Subject access requests to more than 150 organizations were submitted in her name without her direct participation or interaction. When organizations requested additional information to complete these requests, only a small subset of publicly-available data was considered available to the attacker. The ultimate responses of organizations, and any personal data which they provided, was recorded and analyzed to present a broad overview of GDPR right of access practices. 

This experiment was devised as a cursory assessment of the status quo. Having only a single target, and one known to the attacker, likely introduced some biases in the findings. While efforts were made to reduce these experimental biases (see Section~\ref{sec:threat-model}), future research with several diverse and anonymous victims would represent a methodological improvement over this initial investigation.

\subsection{Threat Model}
\label{sec:threat-model}

Our threat model sought to replicate the capabilities of a highly constrained attacker. Generally, the constraints on our attacker can be grouped into two broad categories: knowledge constraints and operational constraints.

\subsubsection{Knowledge Constraints}
\begin{table}
\caption{OSINT Attacker Knowledge}
\label{tab:osint-knowledge}
\resizebox{\linewidth}{!}{%
\begin{tabular}{@{}ll@{}}
\toprule
\textbf{Starting Information} & \textbf{Possible Source} \\ \midrule
Full Name & LinkedIn \\
Free Emails & \begin{tabular}[c]{@{}l@{}}Guessing \\ (e.g. first.last@gmail.com)\end{tabular} \\
Professional Emails & Company Website \\
Phone Numbers & Personal Website \\ \bottomrule
\end{tabular}%
}
\end{table}

The knowledge constraints imposed on our attacker revolved around the information about the victim which the attacker was permitted to incorporate into their social engineering attacks. Specifically, our attacker was assumed to know nothing other than basic public information about the target derived from informed guesswork and open-source intelligence (OSINT). This information is summarized in Table~\ref{tab:osint-knowledge}. The OSINT restriction ensured that our simulated attacker represented the weakest possible form of the attack. Further research on the capabilities of more targeted or informed attackers (e.g. an attacker who has stolen someone's wallet) may prove valuable.

As the attack progressed, the attacker was permitted to use any information they found about the victim to supplement this OSINT knowledge base. So, for example, if a targeted organization provided the attacker with a home address for the victim, the attacker was allowed to include that home address in future requests.

\subsubsection{Operational Constraints}
\label{sec:operational-constraints}
The operational constraints on the attacker revolved around the actions which the attacker could take in managing requests. Again, we sought to replicate a weak form of the attack, leaving open opportunities for research on more capable attackers. Specifically, the attacker was considered incapable of doing anything other than sending emails and falsifying simple documents (e.g. postmarked envelopes). Other capabilities, such as the ability to forge signatures and identity documents, or the ability to spoof email headers, were not assessed in this experiment. 

Additionally, due to legal concerns, no falsified documents were created for the experiment, but instead legitimate documents were submitted in such a way as to replicate the capabilities of the attacker (e.g. a genuine bank statement with all account information covered up). We would expect future work which replicates a more sophisticated attacker - either via technical means (such as email account hijacking or spoofing) or via physical means (such as passport forgery) - would likely have substantially higher success rates than this baseline threat model.

\subsection{Implementation and Delivery}
One of the principle goals of the attack was to develop a mechanism for social engineering at scale. By targeting a large number of organizations, even if only a minority of them prove vulnerable, that proportion may be sufficient for the attacker's purposes. As such, a generic and deliberately vague GDPR subject access request letter was devised which would apply to almost any imaginable organizations but would still appear legitimate. This letter was then templatized and sent to more than 150 organizations using a simple python mailer script.

\subsubsection{SAR Design Considerations}
In this section, we will point out a few aspects of the letter which were designed with the explicit goal of increasing the likelihood of attack success. The full text of the subject access request letter appears in Appendix A. 

One of the principle benefits of targeting GDPR processes for social engineers is the severe time pressures which the law imposes on organizations to issue a response. Given that an organization typically has only one calendar month to respond to any given GDPR request (with limited extensions of up to one additional month), broad or complicated GDPR requests can be difficult to respond to within the allotted timeframe. Under these pressure dynamics, we hypothesized that organizations may be tempted to take shortcuts or be distracted by the scope and complexity of a request and pay less attention to the identity verification aspects of the law. 

As such, the letter is deliberately vague with regards to the requested data, asking for ``any personally identifiable information that [the] organization (or a third party organization on [the organization's] behalf) stores.'' To further compound this complexity, the letter requests not only digital account information, but also data located in ``physical files, backups, emails, voice recordings, or other media.'' This vagueness permits the letter to apply in a wide range of circumstances without organization-specific tailoring.

Additional complexity is cultivated through two requests for information that are not directly relevant to the attacker's objectives. First, the letter requests information about any data-sharing relationship with third parties - requiring an organization to not only identify information they hold about the data-subject but also the origin of that data. A response containing this information has the added benefit of identifying additional targets for future attacks against the victim. 

Second, the letter requests information regarding if personal data has "been disclosed inadvertently [...] as a result of a security or privacy breach." This query is something of a red-herring and is added to suggest a plausible purpose for the subject access request. The intention is to cause a receiving organization to suspect that the attacker has some knowledge of a disclosed or undisclosed data breach and is using the GDPR request to establish grounds for a lawsuit or regulatory complaint. We hypothesized that such a belief might cause organizations to overlook identity verification abnormalities.

Beyond adding complexity, the letter also attempts to preempt one of the principle causes of attack failure - a request for proof of identity. The letter does this by offer to provide identity documents which are ``proportional'' to the data subject's existing relationship with an organization. This interpretation of GDPR is mildly supported by the law itself, which states that data controllers ``shoud not retain personal data for the sole purpose of being able to react to potential requests'' and that data controllers should use ``all reasonable measures to verify the identity of a data subject''~(Rec. 64)\cite{noauthor_general_2016}. Proportionality is an ambiguous standard which may be difficult to determine in the case of many organizations which have no direct relationship with a consumer (such as advertising data-brokers). This gives the attacker justification later on to credibly refuse to provide certain forms of identity documents.

To further bolster this preemption, the letter also specifies that the attacker is only willing to provide identity documents through a ``secure, online portal.'' This unwillingness to provide proof of identity via e-mail can be further supported through referencing guidance on the UK information commissioner's website suggesting that e-mail is a risky means of transferring personal data~\cite{informationcommissionersofficeukSendingPersonalData2018}. 

This demand leaves an organization with three sub-optimal choices. First, they can provide the secure, online portal as requested. However, if such a portal does not exist - as is often the case for small and mid-size businesses - they only have a single calendar month to acquire such a capability. Second, they can refuse to provide a portal and insist that documents be sent over e-mail. This means that they are potentially denying fundamental rights to a data-subject unless the data-subject is willing to take an unnecessary risk with their data. In the event of a subsequent breach of the organization's mail servers, this behavior may be regarded unfavorably by regulators. Finally, the organization can elect not to request identity documents at all, or to request weak forms of identity a user is comfortable sending via email. In our case, this is an ideal outcome for the attacker.

\subsubsection{Delivery and Targeting}
The subject access request letter was designed as a modular .pdf template in which both the victim's details and the target organization's name could be dynamically altered. A fake email account impersonating the victim, of the format: \textit{[first\_name][middle\_initial][last\_name]@gmail.com} was created to send the malicious letters. A simple python script was used to deliver the attack to 150 organizations in two waves of 75 organizations each. Letters in the second wave were bolstered with additional information about the victim that was acquired in the first wave.

No particularly rigorous methodology was employed to select organizations for this preliminary study. We attempted to replicate an attacker who had no prior knowledge of the victim and, as such, selected a handful of well-known organizations within various sectors (e.g. travel or retail). Upon completion of the experiment, we determined that slightly more than half of the queried organizations actually held personal data on the target. Our findings are likely heavily biased towards organizations which do business in the United States and United Kingdom - the two countries with which the authors are most familiar. More rigorous future research, especially research which considers regional and linguistic variation, is likely warranted. However, even this crude initial approach was sufficient to provide weak quantitative insights into the nature of various organizational responses to malicious subject access requests.

\subsection{Ethical and Legal Concerns}
The experiment was designed around a number of ethical and legal constraints. The only subject of the experiment is also a co-author on this paper and, while she was not permitted to participate directly in order to limit bias in the experimental findings, she was kept informed and re-affirmed her consent at all stages of the experiment. Prior to publication of both this paper and the corresponding presentation at Black Hat USA 2019, all screenshots and extracts from responses were redacted to remove sensitive information. These redactions were reviewed and accepted by the experiment's subject.

In engaging with corporations, GDPR requests were submitted with the data subjects' genuine interest in reviewing the personal information which they may store about her. When organizations asked for forms of identification, such as passports or driver's licenses, no attempt to falsify these documents was made - although we suspected that few organizations had the internal capacity to verify the legitimacy of such documents. Similarly, when phone interviews or sworn statements were required, although these are readily falsified, no attempt to do so was attempted. In a small number of situations where a weak form of identity verification was requested (such as a postmarked envelope), legitimate copies of these documents were provided with the data subject's informed consent to assess what the ultimate impact of their forgery might have been.

With regards to vulnerability disclosure, we have directly contacted and informed some organizations with the most severe findings. Unlike software vulnerabilities, the disclosure of social engineering vulnerabilities may have undesirable secondary effects, such as scapegoating the individual who ``fell'' for the attack even when their doing so was a result of poor organizational policy. As the focus of our research was on the broad systemic impact of privacy laws, rather than an attempt to ``catch-out'' any specific company, we made a case-by-case judgement on the appropriateness of disclosure vis-a-vis the severity of our findings. On account of these considerations and to maintain focus on the broader objective of this research, we have elected not to name any specific organizations which were vulnerable to the attacks.

\section{Case Study Results}
Our survey of around 150 organizations revealed significant diversity in the implementation of the subject access request process and, more specifically, within identification standards applied to that process. While we found that many organizations implemented reasonable security controls, we also found many either implemented insecure identity verification controls or no controls at all. Beyond these immediately relevant findings, the case study also revealed many other details which may be useful to researchers, legislators, and organizations interested in the real-world manifestations of GDPR compliance.
\begin{figure}
    \includegraphics[width=\linewidth]{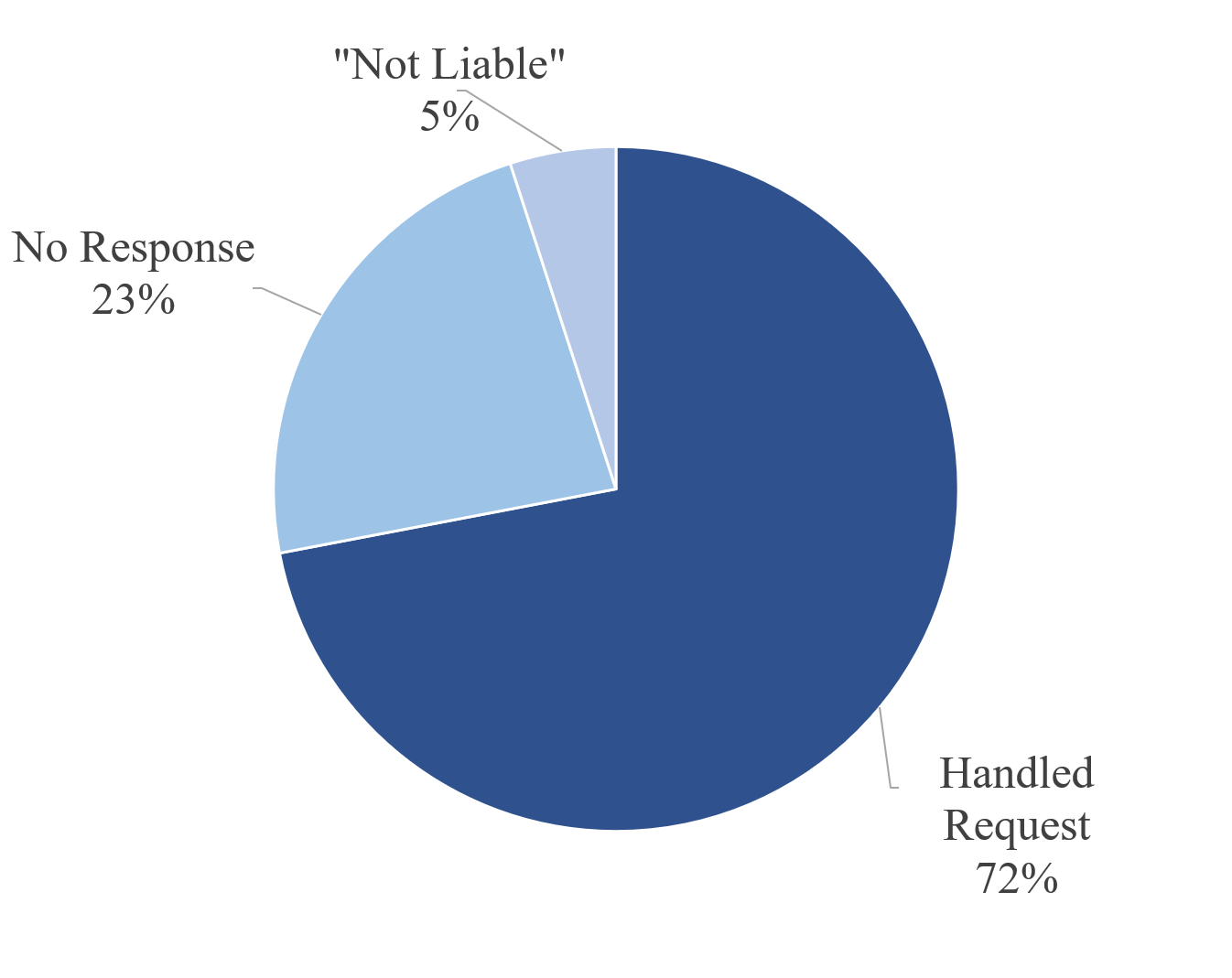}
	\caption{The initial responses of organizations who received a malicious subject access request.}
	\label{fig:response-rates}
\end{figure}

\subsection{Initial Responses}
Not all of the organizations contacted had GDPR subject access request processes in place (see Figure~\ref{fig:response-rates}). Approximately one quarter of organizations contacted never responded to the request. It is unknown if this is because they somehow determined it was illegitimate or if they simply did not have a process in place to respond to subject access requests submitted under GDPR. Within this quarter, a small number of companies responded by saying that they believed GDPR did not apply to them due to jurisdictional constraints. This response came not just from small businesses but also four large (Fortune 250) companies doing business primarily in the United States. These organizations contended that, even though the data subject was a European resident, she had no rights to review the data they held about her under GDPR due to their nature as American businesses.

Of those organizations which did respond to GDPR requests, approximately two thirds responded in such a way as to reveal whether or not the victim had used their services. While the severity of account enumeration can vary depending on the specific circumstance, in some cases the mere existence of an account can reveal deeply sensitive information about an individual, such as in the case of the Ashely Madison data breach in 2015~\cite{kuchlerAshleyMadisonAgrees2016}. In our experimental findings, online dating services were among those organizations which enumerated the presence of user accounts in response to GDPR requests.

\subsection{Ultimate Responses}
\begin{figure*}
    \includegraphics[width=\textwidth]{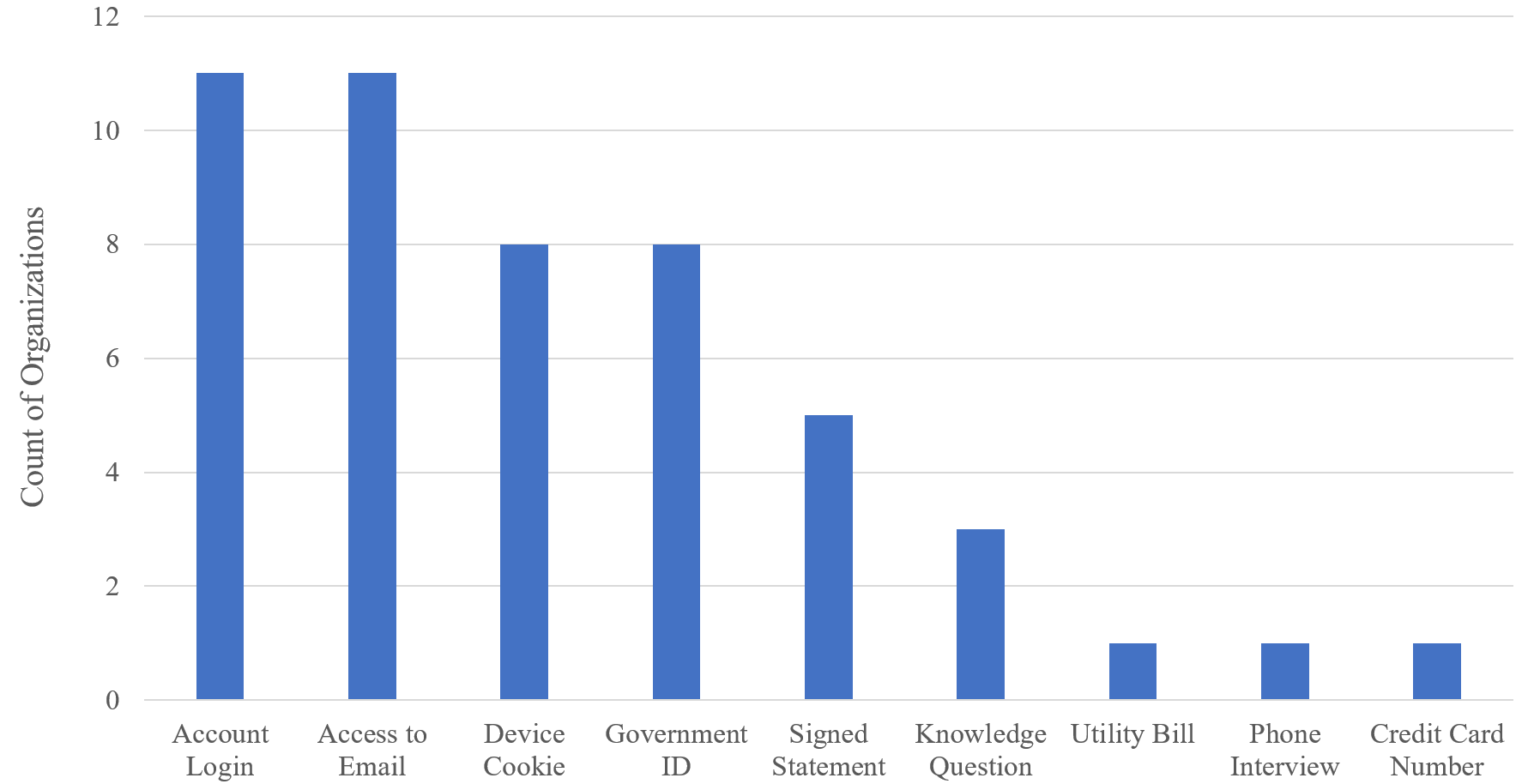}
	\caption{The types of identity requested by organizations.}
	\label{fig:id-types}
\end{figure*}
\begin{figure}
    \includegraphics[width=\linewidth]{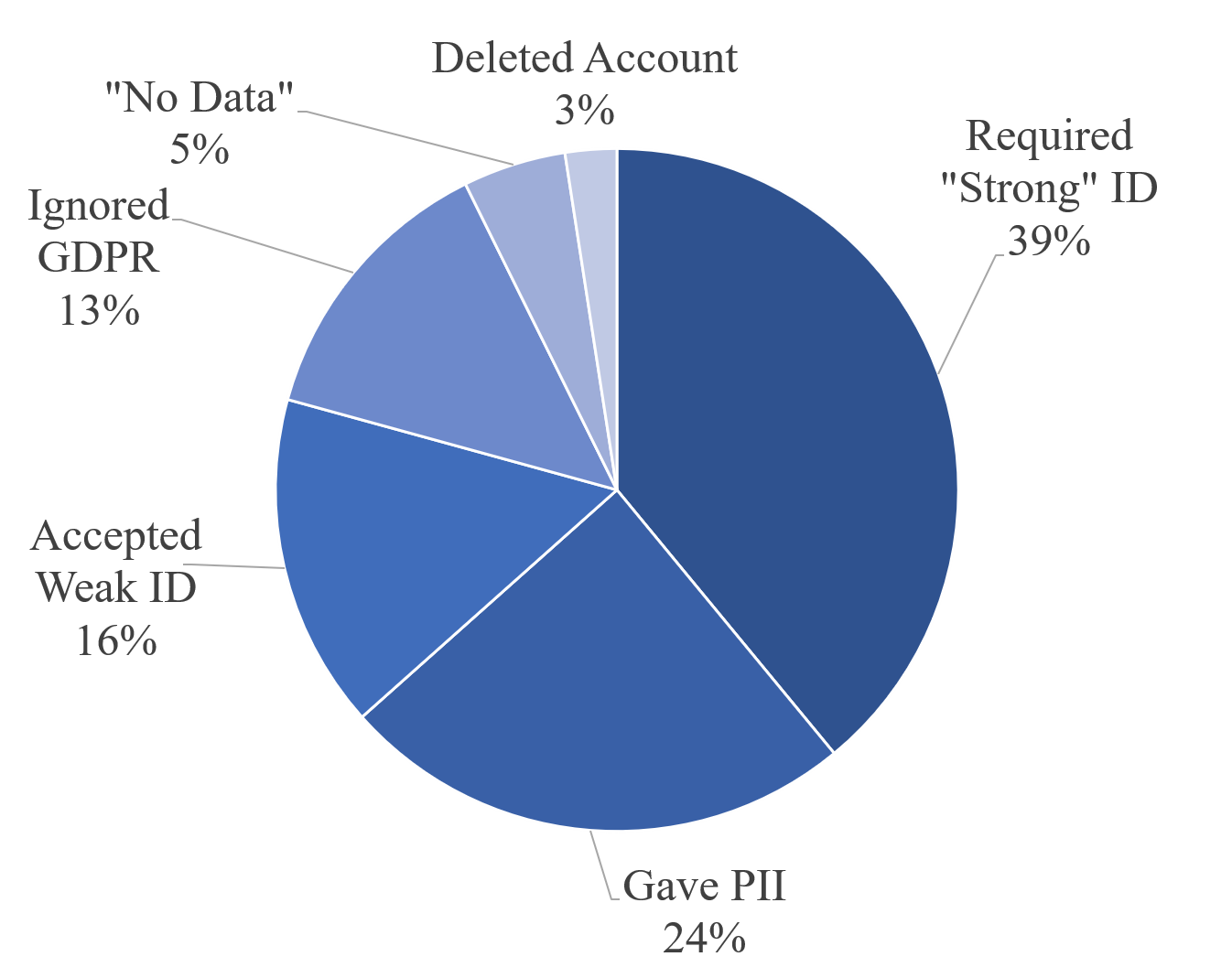}
	\caption{The ultimate responses of organizations who received a malicious subject access request.}
	\label{fig:reveal-rates}
\end{figure}

Within the subset of organizations which had information about our simulated victim, around a quarter provided sensitive information without verifying the identity of the requester (Figure~\ref{fig:reveal-rates}). A further 15\% of organizations contacted requested a form of identity that we believed could easily be stolen or forged (such as a device identifier or a signed statement swearing to be the data subject) but which we did not attempt to falsify (see Section~\ref{sec:operational-constraints}). A small number of organizations (5\%) claimed not to have personal information about the data subject even though she did have an account controlled by the organization. Finally, a handful of organizations (3\%) misinterpreted the subject access request letter as a data removal request and deleted the data subject's account without requiring any further identity verification.

On a more positive note, around 40\% of organizations requested a form of identification which was beyond the reach of our threat model. We observed significant variability in the form of identity requested, which suggests that there is no clear ``best-practice'' (Figure~\ref{fig:id-types}). The most common form of identity verification, and one which met the proportionality test proposed in section~\ref{sec:threat-model}, was to require an email from the original email account used to register with the organization or a login to the data subject's account. 

Not all organizations had access to either of these forms of identity (such as in the case of data-brokers or physical retailers). In these instances, government-issued identity documents, while disproportionate, were the favored mechanism for identity verification. Some organizations used somewhat novel forms of knowledge-based identity verification, perhaps motivated by the proportionality test suggested in the malicious subject access request letter, such as knowledge of the last retail location the data-subject visited or information about the account creation date. This sort of information was beyond the scope of our threat model, but a closer investigation into to feasibility and security of such knowledge-based proofs of identity may prove worthwhile.

With regards to industry-specific tendencies, our sample set was not large enough to make conclusive observations. However, our findings suggest that industries which regularly handle sensitive information (e.g. airlines or banks) or which frequently receive GDPR requests (e.g. social media organizations or consumer-tech giants) tended to be less vulnerable to this specific attack. Meanwhile, organizations in more esoteric sectors (such as education or entertainment) with less experience handling sensitive identity documents or processing GDPR requests tended to be more likely to reveal sensitive information. We found that the largest organizations in our data set (e.g. Fortune 100 companies) tended to perform well and that the smallest organizations tended to simply ignore GDPR requests. Non-profits and mid-size organizations (100 - 1,000 employees) accounted for around 70\% of mishandled requests. This may suggest that there is a ``social-engineering sweet-spot'' targeting organizations large enough to be aware of and concerned about GDPR, but also small enough to not have dedicated significant resources towards compliance. A broader survey capable of robustly validating these initial trends is likely merited.

\subsection{Extended Engagements}
In a handful of instances where an organization requested a strong form of identity verification, we found that these requirements were flexible in practice. For example, after telling a large online gaming company (more than \$1bn in annual revenues), that we had forgotten our account password and could not log in to verify our identity, the organization simply provided the data they held without any further verification.

\begin{figure}[h]
    \includegraphics[width=\linewidth]{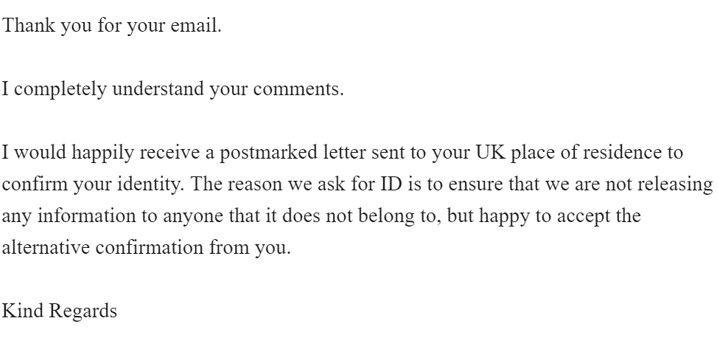}
	\caption{A screenshot of a response which agreed to take a postmarked envelop in lieu of a passport photocopy as proof of identity.}
	\label{fig:postmarked-envelope}
\end{figure}

In other instances, it was not possible to completely eliminate the identity verification process but it was possible to substantially weaken it. For example, a major rail services operator in the United Kingdom initially requested a passport photocopy as proof of identity but, after some negotiation, agreed to accept a postmarked envelope instead (Figure~\ref{fig:postmarked-envelope}. Similarly, several organizations - including a cybersecurity services company - accepted a heavily redacted photograph of a bank statement which contained no banking information beyond the victim's name and address. This suggests that, even when identity documents are requested, little effort is made to verify their authenticity.

\subsection{Data Findings}
In total, over 60 distinct instances of personal information were acquired in the experiment. We defined an ``instance'' as previously unknown personal information of a particular type (e.g. phone numbers) from a given provider. So, for example, if a provider responded with a single previously unknown phone number and a list of 15 previously unknown IP addresses, that would be constitute two ``instances'' of personal information leakage in our analysis.

To better understand these findings, we classified each instance in terms of low, medium, or high severity. These classifications were necessarily arbitrary but were made according to our expectation regarding the exploitability of the information in question. 

Low severity instances (approx.~25\%) consisted of data which did not have obvious direct applications to further attacks against the victim. This included advertising profiles and public records assembled by background-checking agencies. A small portion of account enumeration vulnerabilities relating to sensitive accounts (e.g. online dating profiles) were also included in this category. This data was not intuitively useful in our threat model but may, in certain instances, provide an attacker with insight to help them target further attacks against their target or engage in other social engineering operations (such as blackmail).

Medium sensitivity instances (approx.~60\%) consisted of data which might be of plausible utility to an attacker but only in certain circumstances or with significant additional effort. This included information like standardized test scores, phone numbers, and historical location and purchasing data. Some of this information (e.g. previous residential addresses) could be used to bolster subsequent GDPR requests with additional identifiers. Other information (e.g. detailed purchase histories) might be used to impersonate a credit card provider or bank as a component of a phishing operation. Finally, some of the information provided a deep insight into personal activities and behaviors - such as a complete record of rail journeys taken by the data subject over the past several years or a complete record of hotel stays with a major UK hotel chain.
\begin{figure}[h]
    \includegraphics[width=\linewidth]{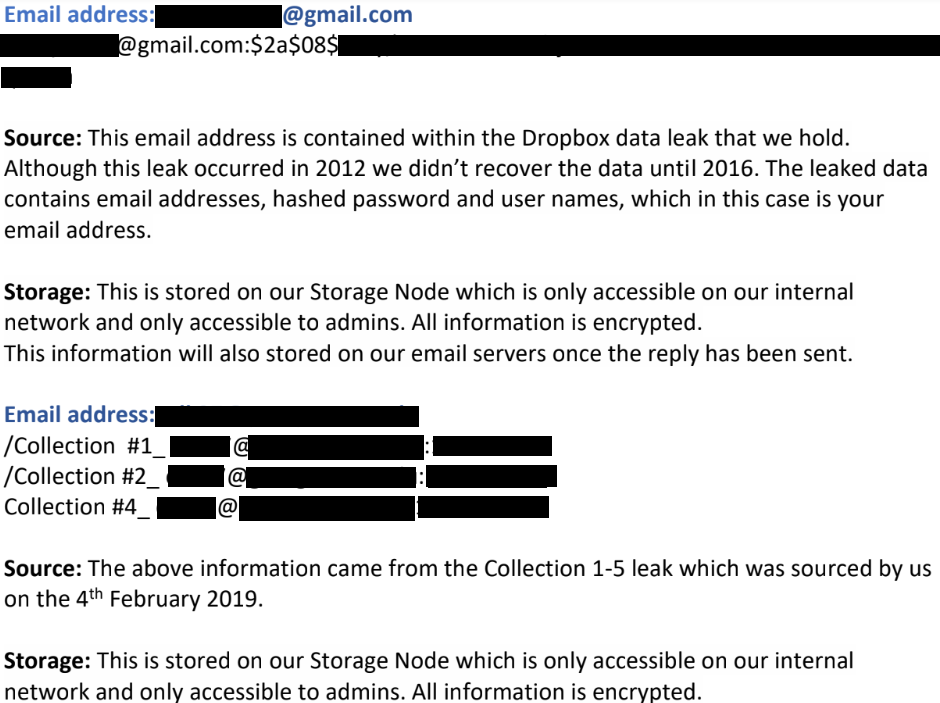}
	\caption{A redacted response from a threat intelligence firm which provided passwords and password hashes from previous (since removed) pastebin credential dumps. The data subject had never heard of or directly engaged with this organization.}
	\label{fig:threat-intel}
\end{figure}

High sensitivity instances (approx.~15\%) consisted of data which was of obvious utility to an adversary. For example, a major education services organization provided the data subject's full US Social Security Number without requesting any identity verification prior to doing so. Based on information on the organization's website, we expect that the records of more than 10 million individuals may have been susceptible to this attack. Similarly, several organizations provided partial information about the victim's credit card numbers. By the end of the attack, 10 digits of the victim's card number, the card's expiration date, originating bank, and postcode were all known to the attacker. While this may not have been sufficient to make actual purchases, it may have been sufficient to impersonate the target's bank in social engineering operations - especially when combined with the medium sensitivity purchase histories discussed previously. In one particularly novel instance, a threat intelligence firm provided the attacker with previously breached usernames and passwords belonging to the data subject~(Figure~\ref{fig:threat-intel}). These passwords were subsequently tested and found working on at least ten online accounts, including an online banking service. 

These findings demonstrate the potential severity of mishandling right of access requests. Unlike other social engineering typologies, such attacks provide readily justified and complete access to an individual's entire data profile with an organization. Moreover, many of the organizations that hold data about a target may be organizations whom the target has never heard of or interacted with directly (such as the aforementioned threat intelligence firm). In such instances, it is nearly impossible for the target to know that their data is stored insecurely or to know that their information may have been compromised.

\section{Proposed Remediations}
There is clear need to address these vulnerabilities. To an extent, this research is a first step in so far as it raises awareness of this attack vector and encourages businesses to think critically about their subject access request process. However, there are also tangible changes which legislators, businesses, and individuals might consider to improve the status quo. We suggest a handful of the most intuitive here but future work considering the optimal approach to identity verification for right of access requests is likely merited.

\subsection{Legislators and Regulators} 
Legislators and regulators are well-suited to attempt to remediate these issues. With minor modifications to GDPR, much needed clarity on reasonable forms of identity verification for data rights requests could be provided. Even without modification to existing legislation, information commissioners could provide tangible guidance to organizations regarding appropriate forms of identification might be for a given right of access request. Such guidance could also be incorporated into future privacy laws modeled on GDPR.

Perhaps even more importantly, legislators can weaken many of the factors which encourage businesses to improperly implement identity verification. Simply assuring businesses that rejecting a suspicious right of access request in good faith will not later result in prosecution if it turns out that the request originated from a legitimate but suspiciously-behaving data subject may be all that's needed for many of the organizations implicated in this study.

In the longer term, legislators may consider offering government-mediated identity verification services for data subjects. Existing services, such at the UK's ``Gov.UK Verify'' service may be scaled up to provide simple yes/no answers on proof of identity to businesses seeking to validate an individual~\cite{gov.ukIntroducingGOVUK2019}. Then, instead of sending sensitive passport documents to a retailer, for example, a consumer could send them to a trusted government service. This would allow for individual data subjects to reap the benefits of strong identity validation without sharing sensitive data with an untrusted third party in order to do so.

\subsection{Businesses}
Absent changes to legislation or improved regulatory clarity, businesses can still attempt to better protect themselves and their customers from this class of attack. 

About 40\% of businesses included in the case study took an approach which was beyond the capabilities of our low-level threat model. For most organizations, this was simply a matter of requiring subject access requests to originate from an email previously known to belong to the data subject or requiring a data subject to log in to their online accounts. If these two identity modes are unavailable, requesting government-issued photo ID is likely the most robust way to prevent this attack. However, organizations who are incapable of adequately protecting this data, or verifying its authenticity, should consider outsourcing these services to a third party.

Businesses should also regularly assess their subject access request process for vulnerabilities and train individual service representatives on detecting and responding to such attacks. Incorporating malicious subject access requests, like the one used in this paper, as a component of regular penetration tests may help mitigate these issues before they become a potential data breach.

\subsection{Consumers}
Individual consumers have little influence over how organizations elect to share their data in response to subject access requests. However, there are some basic actions which may prove useful in mitigating the harms of these attacks.

First, individuals may benefit from considering their data footprint and which organizations may hold sensitive information about them. In cases where this data relationship is not needed (e.g. a service which the individual no longer uses), submitting data removal requests may be prudent to limit the potential avenues for information leakage. In cases where data deletion is not an option, individuals may benefit from enquiring with a given business as to if any past right of access requests have been filed in their name and what the ultimate result of those requests was. This may allow an individual to identify and react to data leakage if it has taken place.

Moreover, individuals should be wary of ``knowledge-based'' authentication as proof that a call or email originates from a given business. For example, in our experiments it was possible to obtain a list of dozens of recent purchases made by an individual and many digits of their credit card number. An attacker could use this information to impersonate a bank representative and potentially do further harm to their victim. As such, even knowledge of extremely granular and esoteric data should be treated with suspicion when used as proof of identity.

\section{Conclusions}
In this paper, we have hypothesized that the GDPR right of access request may be a point of vulnerability to social engineering attacks. Through an experiment encompassing 150 organizations, we demonstrated the real-world viability of such attacks. We find that a large proportion of organizations inadequately verify the originating identity behind right of access requests and that, as a result, deeply sensitive information can be acquired in a repeatable and scalable manner by social engineers. We suggest a number of remediations with a focus on individuals, businesses and lawmakers, to help mitigate these attacks.

This paper represents a preliminary and cursory investigation of how privacy laws might be abused by attackers in novel and unanticipated ways. A great deal of future work - both to verify these initial findings and to characterize other related vulnerabilities - is needed to understand the mechanism by which such vulnerabilities arise in privacy laws and how they may best be anticipated and avoided. The aspiration of all privacy legislation is to ensure a safer and more honest internet. Aligning the implementation of these regulations with their ambition is a vital prerequisite to doing so.

{\footnotesize \bibliographystyle{IEEEtran}
\bibliography{james.bib,casey.bib}}
\flushcolsend
\onecolumn


\section*{Supplementary material}

# Appendix A: Malicious Subject Access Request Letter

<span style="color:red">[TARGET'S ADDRESS]
[ATTACKER'S EMAIL]</span>

Dear Sir or Madam,

**Subject Access Request**

I hope all is well. I am writing to initiate a subject access request under my rights as a European resident and a resident of the United Kingdom. Please supply the information about me I am entitled to under the Data Protection Act 1998 (UK) and the General Data Protection Regulation 2018 (EU) as a data subject and user of [COMPANY NAME] services in a commonly used and machine readable format.

In particular, please supply any personally identifiable information that your organization (or a third party organization on your behalf) stores about me. Please include data that your organization holds about me in your digital or physical files, backups, emails, voice recordings or other media you may store.

If you are additionally collecting personal data about me from any source other than me, please provide me with information about these sources, as referred to in Article 14 of the GDPR.

Finally, I would like to request information regarding if my personal data has been disclosed inadvertently by your company in the past, or as a result of a security or privacy breach.

In addition to my name, this email address, and the postal address indicated above, my accounts may be affiliated with the following identifiers:

- Name: [TARGET FULL NAME]
- Email: [ATTACKER'S EMAIL]
- Email: [FIRSTNAME.LASTNAME@gmail.com]
- Email: [OTHER TARGET EMAILS (e.g. from company website, data breach pastes)]
- Phone: [TARGET PHONE(S)]

If you need any more information from me, or a reasonable fee, please let me know as soon as possible. If you require identity documents to complete these requests, provided that the sensitivity of these documents is proportional to the data I have already consented to allow your organization to store, I am willing to provide these documents via a secure, online portal as soon as possible.

It may be helpful for you to know that a request for information under the GDPR should be responded to within 1 month.

If you do not normally deal with these requests, please pass this letter to your Data Protection Officer. If you need advice on dealing with this request, the Information Commissioner Office can assist you and can be contacted on 0303 123 1113 or at ico.org.uk

Yours faithfully,

-[TARGET NAME]


\label{sec:appendix-letter}
\end{document}